%% file: acit.tex
\magnification=1200
\hsize=16.0truecm
\vsize=24.0truecm
\baselineskip=13pt
\pageno=0
\footline={\ifnum\pageno=0\hfil\else\hss\tenrm\folio\hss\fi}
\input macro-hs

~~~ \hfill CERN-TH/96-172\par
~~~ \hfill BI-TP 96/25
\vskip 2 truecm
\centerline{\bf COLOUR DECONFINEMENT}
\medskip
\centerline{\bf IN HOT AND DENSE MATTER$^*$}
\vskip 1.5 truecm
\centerline{\bf Helmut Satz}
\medskip
\centerline {Fakult\"at f\"ur Physik, Universit\"at Bielefeld, D-33501
Bielefeld, Germany}
\par
\centerline{and}
\par
\centerline{Theory Division, CERN, CH-1211 Geneva 23, Switzerland}
\vskip 2 truecm
\centerline{\bf Summary:}
\medskip
We first introduce the conceptual basis of critical behaviour in
strongly interacting matter, with colour deconfinement as QCD
analog of the insulator-conductor transition and chiral symmetry
restoration as special case of the associated shift in the mass of
the constituents.
Next we summarize quark-gluon plasma formation in
finite temperature lattice QCD. We consider the underlying symmetries
and their spontaneous breaking/re\-sto\-ra\-tion in the transition,
as well as the resulting changes in thermodynamic behaviour.
Finally, we turn to the experimental study of strongly interacting
matter by high energy nuclear collisions, using charmonium
production to probe the confinement status of the produced
primordial medium. Recent results from $Pb$-$Pb$ collisions at CERN
may provide first evidence for colour deconfinement.

\vfill
\noindent \hrule
~~~\par\noindent
* Talk given at CRIS '96, First Catania Relativistic Ion
Studies, Acicastello, Italy, May 27 - 31, 1996; to appear in the
Proceedings.\hfill\break
~~\par
\noindent CERN-TH/96-172
\hfill \break
\noindent BI-TP 96/25
\hfill \break
\noindent June 1996
\eject
\noindent{\bf 1.\ Critical Behaviour in Strongly Interacting Matter}
\medskip
At sufficiently high temperatures or densities, matter is expected to
undergo a transition from confined to deconfined quarks as its
basic constituents. It is, according to quantum chromodynamics (QCD),
always made up of quarks and gluons; in the confined phase, these
coloured constituents are bound to colour-neutral hadrons, while in the
deconfined quark-gluon plasma (QGP), there are freely moving colour
charges.
\par
We begin with a look at the critical behaviour expected during the
transition from hadronic matter to QGP. The Coulomb potential
between two electric charges becomes Debye-screened in a medium of many
other charges, reducing its range to the Debye radius $r_D$,
$$
{e\over r} \to {e \over r}~e^{-r/r_D}. \eqno(1)
$$
At high enough density, when the Debye radius becomes shorter than the
atomic radius, bound electrons are liberated into the conduction band,
changing an insulator into a conductor. Similarly, colour
charges bound by the linearly rising confinement potential of QCD
become screened in a dense medium,
$$
\sigma r \to {\sigma r_c}~[1 - e^{-r/r_c}], \eqno(2)
$$
with $r_c$ as colour screening radius.
At sufficiently high density, colour screening will therefore dissolve a
hadron into its coloured quark constituents, so that deconfinement is
the QCD analog of the insulator-conductor transition.
\par
When the electrons of an insulator are decoupled, their mass is shifted
from the standard $m_e$ to an effective value $m_e^{\rm eff}$,
determined by lattice interactions and the effect of the electron gas
in the conductor. In QCD, we expect the quark mass, which
takes on an effective constituent quark value $m_q^{\rm
const}\simeq m_{\rm proton}/3$ when the quark is confined to a hadron,
to drop back to the current quark value $m_q$ of the QCD Lagrangian once
it is no longer confined.
Such a quark mass shift is therefore another aspect to be considered in
the course of colour deconfinement. In the limit $m_q=0$, the Lagrangian
becomes chirally symmetric, so that in this case the intrinsic chiral
symmetry of the theory must be spontaneously broken in the hadronic
phase and restored in the QGP.
\par
The basic condition for the transition from hadronic to quark matter is
a sufficiently high density of constituents: it then becomes impossible
to define a given quark-antiquark pair or a quark triplet as some
specific hadron, since within any hadronic volume there are many other
possible partners. Such a density can be achieved \break either by
compressing
baryons (cold nuclear matter) or by heating a mesonic medium, increasing
its density by particle production in collisions (hot mesonic matter).
The phase diagram of QCD can thus maps out regions in the plane of
temperature $T$ and baryochemical potential $\mu$ (see Fig.\ 1), with
the latter specifying the mean baryon number density (baryons minus
antibaryons).
\par
How many phases are there in QCD thermodynamics? As we shall see
shortly, it is
known from lattice QCD studies that for $\mu=0$, deconfinement and the
approximate chiral symmetry restoration associated to light quarks occur
at the same temperature $T_c$. So far, technical reasons prevent us from
carrying out lattice calculations for $\mu \not= 0$, leaving in
particular a ``terra incognita" in the low temperature, high density
region. Since the potential between quarks contains an attractive
component, it is quite conceivable that after deconfinement there will
be diquark formation, with quark pairs playing the role of Cooper pairs
in a diquark phase similar to a superconductor. Only at high enough
density or temperature, such diquarks would then break up to form the
true QGP. It appears that this really interesting question of the low
temperature structure of QCD matter will have to remain unanswered until
a suitable lattice scheme is developed for baryonic matter.
\par\vskip 9truecm
\centerline{Fig.\ 1: The phase structure of strongly interacting matter}
\bigskip
It is obviously of great interest to estimate the hadron-quark transition
temperature, and this can be done in various phenomenological models --
bootstrap model, bag model, string model, dual resonance
model or even percolation theory.
It is quite reassuring that they all lead to very similar values.
Let us consider the perhaps simplest picture by assuming the hadronic
phase to be an ideal gas of massless pions, the quark phase an ideal gas
of massless quarks and gluons, based on colour SU(3) and the two
light quark flavours. The pressure of the former is
$$
P_h = 3~{\pi^2 \over 90}~ T^4 \simeq {1 \over 3}~ T^4, \eqno(3)
$$
taking into account the three charge degrees of freedom of a pion. For
the quark-gluon system we get
$$
P_q = [2\times 8 + {7\over 8}(2^3 \times 3)]~{\pi^2 \over 90}~T^4 -
B \simeq 4~ T^4 - B, \eqno(4)
$$
with two spin and eight colour degrees of freedom for the gluons, two
spin, two flavour, two particle-antiparticle and three colour degrees
for the quarks. The bag presssure $B$ characterizes the difference
between the physical vacuum and the ground state of QCD; it provides a
model for the confining feature of the theory, forcing the quarks into a
hadronic volume. Since the preferred thermodynamic state is that of
highest pressure (lowest free energy), the cross-over point $T_c$
obtained by equating $P_h$ and $P_q$ defines the critical temperature
$$
T_c = \left( {90~ B \over 34~ \pi^2 }\right)^{1/4} \simeq~ 0.7~B^{1/4}
\eqno(5)
$$
separating the low temperature hadron from the high temperature QGP
phase, with a transition which is by construction of first order. Using
a bag pressure value from charmonium spectroscopy ($B^{1/4} \simeq 0.2$
GeV), we find with $T_c \simeq$ 140 MeV the Hagedorn temperature first
obtained through the statistical bootstrap model.
\par
After these phenomenological preliminaries, let us now see what results
about deconfinement can be obtained directly from statistical
mechanics based on QCD as underlying theory.
\bigskip
\noindent{\bf 2.\ Statistical QCD}
\medskip
QCD as the dynamical theory of strong interactions is defined by the
Lagrangian
$$
{\cal L}~=~-{1\over 4}F^a_{\mu\nu}F^{\mu\nu}_a~-~\sum_f{\bar\psi}^f
_\alpha(i \gamma^{\mu}\partial_{\mu} + m_f
-g \gamma^{\mu}A_{\mu})^{\alpha\beta}\psi^f_\beta
~,\eqno(6)
$$
with
$$
F^a_{\mu\nu}~=~(\partial_{\mu}A^a_{\nu}-\partial_{\nu}A^a_{\mu}-
gf^a_{bc}A^b_{\mu}A^c_{\nu})~. \eqno(7)
$$
Here $A^a_{\mu}$ denotes the gluon field of colour $a$ ($a$=1,2,...,8)
and $\psi^f_{\alpha}$ the quark field of colour $\alpha$
($\alpha$=1,2,3) and flavour $f$; the current quark masses are given by
$m_f$. With the dynamics thus fixed, we obtain the corresponding
thermodynamics from the partition function, which is most suitably
expressed as a functional path integral,
$$
Z(T,V) = \int ~dA~d\psi~d{\bar\psi}~
\exp~\left(-\int_V d^3x \int_0^{1/T} d\tau~
{\cal L}(A,\psi,{\bar\psi})~\right), \eqno (8)
$$
since this form involves directly the Lagrangian density defining the
theory.
The spatial integration in the exponent of Eq.\ (8) is performed over
the entire volume $V$ of the system; in the thermodynamic limit it
becomes infinite. The time component $x_0$ is ``rotated"
to become pure imaginary, $\tau = ix_0$, thus turning the Minkowski
manifold, on which the fields $A$ and $\psi$ are originally defined,
into a Euclidean space. The integration over $\tau$ in Eq.\ (8) runs
over a finite slice whose thickness is determined by the temperature
of the system. The finite temperature behaviour of the partition
function in the Euclidean form thus becomes a finite size effect
in the imaginary time direction. Eq.\ (8) is derived from the usual
trace form of the partition function. As a consequence, vector
fields have to be periodic and spinor fields antiperiodic at the
boundaries of the imaginary time integration.
\par
Once $Z(T,V)$ is given, we can
calculate all thermodynamical observables in the usual fashion. Thus
$$
\epsilon = (T^2/V)\left({\partial \ln Z \over \partial T}\right)_V
\eqno (9)
$$
gives us the energy density, and
$$
P = T \left({\partial \ln Z\over \partial V}\right)_T
\eqno(10)
$$
the pressure. What remains is to find a way to actually evaluate these
expressions in the case of a relativistic, interacting quantum
field theory. In QED, there are divergences both for small
(infrared) and for large (ultraviolet) momenta; hence
renormalization is required to get finite results, and it is to be
expected that renormalisation will be necessary for QCD as well.
But there is a further, more serious, problem. The standard evaluation
method for QED -- perturbation theory -- is not applicable to the
study of critical behaviour. Since long range correlations and
multi-particle interactions are of crucial importance here, the
interaction terms cannot be assumed as small. We therefore need a
non-perturbative regularisation scheme for the solution of a
relativistic quantum field theory.  So far, there
is only one method available which fulfills these requirements:
the lattice formulation introduced by K. Wilson \ref{Wilson}.
It puts the thermodynamic observables, such as the energy density (9)
or the pressure (10), into a form that can be evaluated numerically by
computer simulation \ref{Creutz}.
\par
The lattice formulation of statistical QCD is obtained as follows.
First, the continuum integration over space and (imaginary) time in the
action (the exponent in Eq.\ (8)) is replaced by a summation over a
finite space-time grid. To maintain a gauge invariant form, the gluon
fields must be associated to the links between adjacent sites of this
lattice, the quark fields to lattice sites. Next, the resulting
discrete form of the action allows the quark field integration in
Eq.\ (8) to be carried out. This leads to the partition function
$$
Z(T, V) =  \int~\prod_{{\rm links}}~dU
\exp[-S(U)] , \eqno (11)
$$
where the $U$ are unitary matrices formed from the gluon fields.
The temperature is determined by the number $N_t$ of lattice sites in
the imaginary time direction, $T=1/N_t a$, the volume by the
corresponding number in space, $V=(N_s a)^3$, with $a$ denoting
the lattice spacing. The action $S(U)$ in Eq.\ (11) is found to have
the form of a (gauge-invariant) spin system, so that the partition
function becomes structurally equivalent to that of a spin system. And
for the study of such systems, there are known computer simulation
methods.
\par
In computer simulation one essentially creates a lattice world
according to the given dynamics (here QCD) on a large computer
and brings this world into equilibrium by successive Monte Carlo
iterations, using the action $\exp\{-S(U)\}$ as a weight to determine
improved configurations. Once equilibrium is obtained, one measures
any observable of interest on a large number of equilibrium
configurations and thus obtains its value (for a more extensive
survey, see \ref{Ka/La}). In this way, one can
determine the behaviour of the energy density, the specific heat, or
any other desired quantity even in the region of critical behaviour of
the system \ref{Engels}.
\par
However, the method does encounter some problems. We are evidently
interested in true statistical QCD as a continuum theory, not in its
approximation on a discrete lattice. One thus has to study how lattice
results behave in the limit of large lattice size and small lattice
spacing. This extrapolation to the continuum limit requires the study
of lattices of different (large) sizes, so that extensive numerical work
on large-scale computers is needed. Such work has been going on over the
past fifteen years, but for really precise results of the full theory
with quarks, more work is needed. In particular, the smaller the quark
mass is, the more time-consuming the calculations become. Hence present
results still use somewhat too large current quark masses and hence
still obtain in turn somewhat too large a pion mass. The rapid
improvement of computer power and performance supports the expectation
that in the next years, fully realistic calculations will become
feasible. -- A more serious problem is that the method is at present
restricted to studies at vanishing baryon number density. The reason
for this is ``purely technical": for non-vanishing baryochemical
potential $\mu$, the weight $\exp\{-S(U,\mu)\}$ is no longer positive
definite, so that the standard Monte Carlow methods for generating
equilibrium configurations breaks down. It is to be hoped that an
alternative method will be found one of these years. This would then
in particular allow us to address the interesting question of the
phase structure of strongly interacting matter at high density and low
temperature (see Fig.\ 1).
\par
What have we learned so far from the computer simulation of finite
temperature lattice QCD? The first observable to consider is the
deconfinement measure \refs{Larry}{Kuti}
$$
L(T) \sim \exp\{-V(\infty)/T\} \eqno(12)
$$
where $V(r)$ is the potential between a static quark-antiquark pair
separated by a distance $r$. In the limit of infinite current quark
mass, i.e., in pure $SU(3)$ gauge theory, $L(T)$ becomes the order
parameter of a center $Z_3$ symmetry, since $V(\infty)=
\infty$. In the confinement regime, we therefore have $L=0$; colour
screening, on the other hand, makes $V(r)$ finite at large $r$, so that
in the deconfined phase, $L$ does not vanish. In the large quark mass
limit, deconfinement thus corresponds to the spontaneous breaking of
a $Z_3$ symmetry, much
like the onset of spontaneous magnetisation in a spin model. The
structure of the theory here becomes that of a three-state Potts' model,
which shows a first order phase transition. One may thus expect a
similar behaviour for $SU(3)$ gauge theory; for continuous transitions,
one would say that spin and gauge theories are in the same universality
class, and for $SU(2)$ gauge theory and the corresponding $Z_2$ spin
system, the Ising model, this can in fact be shown. For finite current
quark mass $m_q$, $V(r)$ remains finite for $r \to \infty$, since the
string between the two colour charges ``breaks" when the corresponding
potential energy becomes equal to the mass $m_h$ of the lowest hadron;
beyond this point, it becomes energetically more favourable to produce
an additional hadron. Hence now $L$ no longer vanishes in the confined
phase, but only becomes exponentially small there,
$$
L(T) \sim \exp\{-m_h/T\}; \eqno(13)
$$
here $m_h$ is typically of the order of the $\rho$-mass, since the pion
as largely Goldstone boson plays a special role. This gives us $L \sim
10^{-2}$, rather than zero. Deconfinement is thus indeed much like the
insulator-conductor transition, for which the order parameter, the
conductivity $\sigma(T)$, also does not really vanish for $T>0$, but
with $\sigma(T) \sim \exp\{-\Delta~E/T\}$ is only exponentially small,
since thermal ionisation (with ionisation energy $\Delta~E$) produces
a small number of unbound electrons even in the insulator phase.
\par
Fig.\ 2 shows recent lattice results for $L(T)$ and the
corresponding susceptability $\chi_L(T) = \langle L^2 \rangle - \langle
|L| \rangle^2$. The calculations were performed
for the case of two flavours of light quarks, using a
current quark mass about four times larger than that needed for the
physical pion mass \ref{Laermann}.
We note that $L(T)$ undergoes the expected sudden
increase from a small confinement to a much larger deconfinement
value. The sharp peak of $\chi_L(T)$ defines quite well the transition
temperature; gauging the lattice scale in terms of the $\rho$-mass, we
get $T_c \simeq 0.15$ GeV.
\par
The next quantity to consider is the effective quark mass; it is
measured by the expectation value of the corresponding term in the
Lagrangian, $\langle {\bar \psi} \psi \rangle$. In the limit of
vanishing current quark mass, the Lagrangian becomes chirally symmetric
and $\langle {\bar \psi} \psi \rangle(T)$ the corresponding order
parameter. In the confined phase, with effective constituent quark
masses $m_q^{\rm const} \simeq 0.3$ GeV, this chiral symmetry is
spontaneously broken, while in the deconfined phase, at high enough
temperature, we expect its restoration. In the real world, with finite
pion and hence finite current quark mass, this symmetry is also only
approximate, since $\langle {\bar \psi} \psi \rangle(T)$ now never
vanishes at finite $T$.
\par
The behaviour of $\langle {\bar \psi} \psi \rangle(T)$ and the
corresponding susceptability $\chi_m \sim \partial
\langle {\bar \psi} \psi \rangle / \partial m_q$ are shown in Fig.\ 3,
calculated for the same case as above in Fig.\ 2.
We note here the expected sudden drop of the
effective quark mass and the associated sharp peak in the
susceptability. The temperature at which this occurs coincides with that
obtained through the deconfinement measure. We therefore conclude that
at vanishing baryon number density, quark deconfinement and the shift
from constituent to current quark mass coincide.
\par
We thus obtain for $\mu=0$ a rather well defined phase structure,
consisting of a confined phase for $T < T_c$, with $L(T) \simeq 0$ and
$\langle {\bar \psi} \psi \rangle \not= 0$, and a
deconfined phase for $T>T_c$ with $L(T)\not= 0$ and
$\langle {\bar \psi} \psi \rangle \simeq 0$. The
underlying symmetries associated to the critical behaviour at $T=T_c$,
the $Z_3$ symmetry of deconfinement and the chiral symmetry of the quark
mass shift, become exact in the limits $m_q \to \infty$ and $m_q \to
0$, respectively. In the real world, both symmetries become approximate;
nevertheless, we see from Figs.\ 2 and 3, that both associated measures
retain an almost critical behaviour.
\par
Next we turn to the behaviour of energy density $\e$ and pressure $P$ at
deconfinement. The most accurate results exist so far for pure
gauge theory, i.e., for QCD in the infinite quark mass limit; the
qualitative behaviour remains much the same when quarks are included.
We see in Fig.\ 4 that $\e/T^4$ in SU(3) gauge theory \ref{Boyd}
changes quite abruptly
at the above determined critical temperature $T=T_c$, increasing
from a low hadronic value (difficult to determine precisely with the
accuracy of present lattice calculations) to nearly that expected for
an ideal gas of quarks and gluons. The forms shown here are obtained by
extrapolating results from different lattice size calculations to the
continuum limit. Besides the sudden increase at deconfinement, there
are two further points to note. The interaction measure $(\e-3P)/T^4$,
if correctly normalized,
vanishes for an ideal gas of massless particles. It certainly does not
vanish here, as shown more explicitly in Fig.\ 5. In particular, it is
quite strongly peaked in the region $T_c < T < 2~T_c$, indicating the
presence of strong remaining interaction effects in this region. The
nature of these effects is presently one of the main questions in finite
temperature lattice QCD; there are a number of model proposals to
account for the ``measured" lattice data. The second, somewhat
surprising effect is that the thermodynamic observables do not fully
attain their Stefan-Boltzmann values (marked ``SB" in Fig.\ 4)
even at very high temperatures,
in contrast to earlier conclusions based on less
precise calculations. The remaining 10 to 15 \% deviation could well be
due to effective ``thermal" masses of gluons (and of quarks in full
QCD); this problem also is in under investigation.
\par
Finally we turn to the value of the transition temperature. Since QCD
(in the limit of massless quarks) does not contain any dimensional
parameters, $T_c$ can only be obtained in physical units by expressing
it in terms of some other known observable which can also be calculated
on the lattice, such as the $\rho$-mass, the proton mass, or the string
tension. In the continuum limit, all different ways should lead to the
same result. Within the present accuracy, they define the uncertainty
so far still inherent in the lattice evaluation of QCD. Using the
$\rho$-mass to fix the scale leads to $T_c\simeq 0.15$ GeV, while
the string tension still allows values as large as $T_c \simeq 0.20$
GeV. This means that energy densities of some 1 - 2 GeV/fm$^3$ are
needed in order to produce a medium of deconfined quarks and gluons.
\par
In summary, finite temperature lattice QCD shows
\par
\item{--}{that there is a deconfinement transition with an associated
shift in the effective quark mass at $T_c \simeq$ 0.15 - 0.20 GeV;}
\item{--}{this transition is accompanied by a sudden increase in the
energy density  (``latent heat of deconfinement") from a small hadronic
value to a much larger value some ten to twenty percent below that of
ideal quark-gluon plasma;}
\item{--}{for $T_c \leq T \leq 2~T_c$, the ideal gas measure $(\e -
3~P)/T^4$ differs very much from zero, indicating the presence of
considerable plasma interactions.}
\par\noindent
Both conceptually and through {\sl ab initio} QCD calculations we thus
have a relatively good understanding of the critical behaviour
and the phase structure expected for strongly interacting matter.
\bigskip
\noindent
{\bf 3.\  Colour Deconfinement in Nuclear Collisions}
\medskip
How can these predictions be tested? Since ten years, experiments
studying the collision of nuclei at high energies are carried out with
the aim of producing strongly interacting matter in the laboratory.
There are today promising indications that the hadrons produced in such
collisions indeed come from equilibrated thermal systems.
In the next section, we want to address in particular the question of
how one can check if these systems in their early hot and dense stages
consisted of deconfined quarks and gluons. In case of thermal
evolution, the hadrons formed in the final stage do not carry any
information about the earlier stages of the system. We thus need to
find a probe which somehow retains this primordial information.
The most promising and most intensively studied probe of this kind is
the behaviour of charmonium production in nuclear collisions,
proposed ten years ago \ref{Matsui}. With the help of this probe, very
recent experimental results may provide the first indication of the
onset of colour deconfinement.
\par
A hadron placed into a deconfining medium will dissolve into its quark
constituents. If the medium expands, cools off and eventually
hadronizes, normal hadrons will now reappear and, in case of an
expansion in thermal equilibrium, not carry any information about the
earlier stages.
\par
The situation is quite different for a \J~put into a quark-gluon plasma.
The \J~is a bound state of the heavy $c$ and $\bar c$ quark, which each
have a mass of about 1.5 GeV. The J~has a mass of about 3.1 GeV; with a
radius of about 0.3 fm it is much smaller than the usual light hadrons.
It has a binding energy (the difference between \J~mass and open charm
threshold) of about 0.64 GeV, which is much larger than the typical
hadronic scale $\Lambda_{\rm QCD}\simeq 0.2$ GeV. The \J~is produced
quite rarely in hadronic collisions -- at present energies, in about
one out of 10$^5$ events. Through its decay into dimuons,
it is, however, rather easily detectable in suitably triggered
experiments, so that in the ongoing heavy ion studies at CERN
\ref{NA38}, some
hundred thousand \J's are measured for a given target-projectile
combination.
\par
If the quark-gluon plasma is sufficiently hot, a \J~will also melt in
it. However, its constituents, the $c$ and the $\bar{c}$, now separate
and never meet again. Since the production of more than one $\C$ pair
per collision is very strongly excluded, the $c$ must at hadronisation
combine with a normal antiquark, the $\bar{c}$ with a normal quark,
leading to a $D$ and a $\bar{D}$, respectively. If nuclear collisions
produce a deconfining medium, then such collisions must also lead
to a suppression of \J~production \ref{Matsui}.
\par
Before we can use this as a deconfinement probe, we must know if there
are possible interactions in a {\sl confined} medium which can cause
\J~suppression. This question is now answered, including in particular
also a direct experimental test of the theoretical answer. The cross
section for the dissociation of a \J~colliding with a usual light hadron
can be calculated in short distance QCD \ref{KS3}; the relevant diagram
is shown in Fig.\ 6. \par
\vskip 9truecm
\centerline{Fig.\ 6: \J~interaction with a light hadron}
\bigskip
It contains two parts: the break-up of a \J~by an incident
gluon, essentially the photo-effect analog in QCD, and the emission or
absorption of a gluon by a light hadron. The latter is evidently of
non-perturbative nature, but the needed gluon distribution function is
determined in deep inelastic scattering. Starting from the operator
product expansion, one thus establishes sum rules relating the
dissociation cross section $\sigma_{h-\psi}^{\rm in}$ to the gluon
distribution function $g_h(x)$ for a light hadron. The large binding
energy of the \J~requires hard gluons for both resolution and break-up.
On the other hand, the presence of hard gluons in light hadrons of
present momenta is strongly suppressed; $g_h(x)$ falls rapidly for
large $x$. As a result, $\sigma_{h-\psi}^{\rm in}(s)$ suffers a very
strong threshold damping; it is essentially zero until the $h-\psi$
collision energy $\sqrt s$ becomes much larger than the presently
available 20 GeV (Fig.\ 7).
\par
We thus conclude that in the presently produced media, collisions
with hadrons cannot dissociate \J's. This makes \J~suppression into an
unambiguous deconfinement probe: \J's can be suppressed in a given
medium if and only if this medium contains deconfined gluons.
\par
Since the predicted threshold suppression of $\sigma_{h-\psi}^{\rm
in}(s)$ is crucial in this argument, it should certainly be tested
experimentally. So far, it has been confirmed for the very similar
process of \J~photo-production. A direct check is possible, however,
through an ``inverse kinematics" experiment, shooting a heavy nuclear
beam at a hydrogen or deuterium target \ref{KS5}. Such an experiment
should certainly be carried out, and preparations are in progress.
\par
Before we can actually analyse the confinement status of the media
formed in nuclear collisions, one further problem has to be addressed.
We have so
far implicitly considered the fate of a fully formed physical \J~in the
given medium. However, the same collision that produces the medium also
has to produce the \J, and neither production process is instantaneous.
The mechanism of \J~production in hadronic collisions has recently been
established more precisely, both experimentally \ref{FNAL} and
theoretically \refs{Braaten}{KS6}.
The first stage of the production process is the formation of a coloured
$\C$ pair, which combines with a colliner gluon to form a colour-neutral
$\C-g$ state. This pre-resonance charmonium state turns into a physical
\J, i.e., a colour singlet $\C$ state, after some 0.2 to 0.3 fm. In its
pre-resonance stage, however, it can interact with nuclear matter, and
this interaction must be taken into account before studying the
suppression of physical \J's. Theoretical estimates \ref{KS6} give the
$\C-g$
state in interactions with hadrons a dissociation cross section of some
6 - 7 mb. This can be studied in $p-A$ interactions, and recent
precision data \refs{Gonin}{Carlos} confirm the picture both
qualitatively and quantitatively.
\par
The survival probability of the pre-resonance
$\C-g$ state passing through a length $L$ of normal nuclear matter can
be estimated by $S_{\j}(L)=\exp\{-n_0 \sigma_{\C g} L\}$; this estimate
can be checked by calculations based on Glauber theory. Looking at
the survival probability in $p-A$ collisions as function of
$L\simeq (3/4)R_A^{1/3}$ leads
to a cross section $\sigma_{\C-g} = 6.3 \pm 1.2$ mb for the break-up of
pre-resonance charmonia by collisions with nucleons (Fig.\ 8), which
agrees with the theoretical expectations.
\par
We thus find that \J~suppression in $p-A$ collisions is well
described in terms of pre-resonance dissociation. The equality of \J~and
\P~suppression in such interactions moreover finds a natural explanation
by such a process. Turning now to nuclear collisions and the question
of deconfinement, we have to check if such reactions lead to a
suppression beyond the known $\C-g$ dissociation in normal nuclear
matter. The relevant data from $O-Cu$, $O-U$ and $S-U$ interactions at
th CERN-SPS are included in Fig.\ 8; we see that they agree completely
with the predicted pre-resonance suppression. This result can also be
extended to $S-U$ collisions at different impact parameters.
Hence we conclude that up to central $S-U$ collisions, the \J~does not
suffer any nuclear effect beyond the mentioned pre-resonance suppression
in nuclear matter. In other words, these collisions do not produce a
deconfined medium, even though the associated average energy density
is expected to be in the range of 1 - 3 GeV/fm$^3$.
\par
A quark-gluon plasma must thus lead to more \J~suppression than that
obtained from the pre-resonance $\C-g$ absorption. The recent
announcement \refs{Gonin}{Carlos} of a strong ``anomalous"
\J~suppression in $Pb$-$Pb$
collisions at CERN may therefore provide a first hint of deconfinement.
Although the average energy density in $Pb$-$Pb$ collisions is only
slightly higher than that in central $S$-$U$ interactions, the NA50
collaboration finds, as function of collision centrality, a very rapid
onset of much stronger suppression. In particular, central
$Pb$-$Pb$ collisions result in a suppression which is more than twice
that due to pre-resonance absorption in normal nuclear matter, as shown
in Fig.\ 9. The path length $L$ used as a variable there provides
in nucleus-nucleus collisions also a measure of the average energy
density $\e$; hence the anomalous suppression sets in suddenly at a
certain value of $\e$.
\par
A simple model can illustrate how such an effect could arise
\refs{Gupta}{Blaizot}. Although
the average energy densities in central $S$-$U$ and central $Pb$-$Pb$
collisions are very similar, the corresponding energy density profiles
are quite different (Fig.\ 10); the interior of the interaction region
is much hotter in $Pb$-$Pb$ than in $S$-$U$ collisions. If
the peak value in central $S$-$U$ collisions is just the
critical energy density needed for \J~melting, then all \J's formed in
the $Pb$-$Pb$ region hotter than this will be suppressed. It
turns out that this in fact gives just the right amount of anomalous
\J~suppression \ref{Blaizot}.
\par
Clearly more studies, both experimental and theoretical, are needed
before a definite conclusion is possible. At this time, however, it does
seem that if one can confirm
\item{--}{the sudden onset of the anomalous suppression, e.g.\ by more
peripheral $Pb$-$Pb$ collisions, different $A$-$B$ combinations,
different collision energies, and}
\item{--}{the transparency of confined matter to \J's by an
inverse kinematics experiment,}
\par\noindent
then the observed effect would seem to be the first indication of colour
deconfinement in nuclear collisions.
\bigskip
\centerline
{\bf Acknowledgements}
It is a pleasure to thank F.\ Karsch, D.\ Kharzeev, E.\ Laermann and
C.\ Louren\c co for many helpful discussions.
\vfill\eject
\centerline{\bf References:}
\bigskip
\item{\reftag{Wilson})}{K.\ G.\ Wilson, \PR D 10 (1974) 2445.}
\par
\item{\reftag{Creutz})}{M.\ Creutz, \PR D 21 (1980) 2308.}
\par
\item{\reftag{Ka/La})}{F.\ Karsch and E.\ Laermann, Rep. Prog. Phys. 56
(1993) 1347.}
\par
\item{\reftag{Engels})}{J.\ Engels et al., \PL 101 B (1981) 89 and \NP B
205 [FS5] (1982) 545.}
\par
\item{\reftag{Larry})}{L.\ D.\ McLerran and B.\ Svetitsky, \PL 98 B
(1981) 195 and \PR D 24 (1981) 450.}
\par
\item{\reftag{Kuti})}{J.\ Kuti, J.\ Pol\'onyi and K.\ Szlach\'anyi,
\PL 98B (1981) 199.}
\par
\item{\reftag{Laermann})}{F.\ Karsch and E.\ Laermann, \PR D 50 (1994)
6954;\hfill\break
E.\ Laermann, Report at the XXIIth
International Conference on Ultrarelativistic Nucleus-Nucleus
Collisions, Heidelberg/Germany, May 1996.}
\par
\item{\reftag{Boyd})}{G.\ Boyd et al., \PRL 75 (1995) 4169;\hfill\break
G.\ Boyd et al., ``Thermodynamics of SU(3) Lattice
Gauge Theory", Bielefeld Preprint BI-TP 96/04, January 1996, to appear
in \NP B.}
\par
\item{\reftag{Matsui})}{T.\ Matsui and H.\ Satz, \PL B 178 (1986) 416.}
\par
\item{\reftag{NA38})}{C.\ Baglin et al., \PL B220 (1989) 471; B251
(1990) 465, 472; B225 (1991) 459.}
\par
\item{\reftag{KS3})}{D.\ Kharzeev and H.\ Satz, \PL B 334 (1994) 155.}
\par
\item{\reftag{KS5})}{D.\ Kharzeev and H.\ Satz, \PL B 356 (1995) 365.}
\par
\item{\reftag{FNAL})}{See e.g. V.\ Papadimitriou (CDF), ``Production
of Heavy Quark States at CDF", Preprint Fermilab-Conf-95-128 E,
March 1995; \hfill \break
L.\ Markosky (D0), ``Measurements of Heavy Quark Production at D0",
Preprint Fermilab-Conf-95-137 E, May 1995.}
\par
\item{\reftag{Braaten})}{E.\ Braaten and S.\ Fleming, \PRL 74 (1995)
3327.}
\par
\item{\reftag{KS6})}{D.\ Kharzeev and H.\ Satz, \PL B 366 (1996) 316.}
\par
\item{\reftag{Gonin})}{M.\ Gonin (NA50), Report at the XXIIth
International Conference on Ultrarelativistic Nucleus-Nucleus
Collisions, Heidelberg/Germany, May 1996.}
\par
\item{\reftag{Carlos})}{C.\ Louren\c co (NA50), Report at the XXIIth
International Conference on Ultrarelativistic Nucleus-Nucleus
Collisions, Heidelberg/Germany, May 1996.}
\par
\item{\reftag{Gupta})}{S.\ Gupta and H.\ Satz, \PL B 283 (1992) 439.}
\par
\item{\reftag{Blaizot})}{J.-P.\ Blaizot and J.-Y.\ Ollitrault,
``\J~Suppression in Pb-Pb Collisions: A Hint of Quark-Gluon Plasma
Production?", Saclay Preprint T96/039, May 1996.}
\vfill\eject
\vfill\eject\bye

%% file: macro-hs.tex
\def\J{$J/\psi$}
\def\j{J/\psi}
\def\P{$\psi'$}

\def\C{c{\bar c}}

\def\e{\epsilon}

\def\lsim{\raise0.3ex\hbox{$<$\kern-0.75em\raise-1.1ex\hbox{$\sim$}}}
\def\gsim{\raise0.3ex\hbox{$>$\kern-0.75em\raise-1.1ex\hbox{$\sim$}}}

\newcount\REFERENCENUMBER\REFERENCENUMBER=0
\def\REF#1{\expandafter\ifx\csname RF#1\endcsname\relax
               \global\advance\REFERENCENUMBER by 1
               \expandafter\xdef\csname RF#1\endcsname
                   {\the\REFERENCENUMBER}\fi}
\def\reftag#1{\expandafter\ifx\csname RF#1\endcsname\relax
               \global\advance\REFERENCENUMBER by 1
               \expandafter\xdef\csname RF#1\endcsname
                      {\the\REFERENCENUMBER}\fi
             \csname RF#1\endcsname\relax}
\def\ref#1{\expandafter\ifx\csname RF#1\endcsname\relax
               \global\advance\REFERENCENUMBER by 1
               \expandafter\xdef\csname RF#1\endcsname
                      {\the\REFERENCENUMBER}\fi
             [\csname RF#1\endcsname]\relax}
\def\refto#1#2{\expandafter\ifx\csname RF#1\endcsname\relax
               \global\advance\REFERENCENUMBER by 1
               \expandafter\xdef\csname RF#1\endcsname
                      {\the\REFERENCENUMBER}\fi
           \expandafter\ifx\csname RF#2\endcsname\relax
               \global\advance\REFERENCENUMBER by 1
               \expandafter\xdef\csname RF#2\endcsname
                      {\the\REFERENCENUMBER}\fi
             [\csname RF#1\endcsname--\csname RF#2\endcsname]\relax}
\def\refs#1#2{\expandafter\ifx\csname RF#1\endcsname\relax
               \global\advance\REFERENCENUMBER by 1
               \expandafter\xdef\csname RF#1\endcsname
                      {\the\REFERENCENUMBER}\fi
           \expandafter\ifx\csname RF#2\endcsname\relax
               \global\advance\REFERENCENUMBER by 1
               \expandafter\xdef\csname RF#2\endcsname
                      {\the\REFERENCENUMBER}\fi
            [\csname RF#1\endcsname,\csname RF#2\endcsname]\relax}
\def\refss#1#2#3{\expandafter\ifx\csname RF#1\endcsname\relax
               \global\advance\REFERENCENUMBER by 1
               \expandafter\xdef\csname RF#1\endcsname
                      {\the\REFERENCENUMBER}\fi
           \expandafter\ifx\csname RF#2\endcsname\relax
               \global\advance\REFERENCENUMBER by 1
               \expandafter\xdef\csname RF#2\endcsname
                      {\the\REFERENCENUMBER}\fi
           \expandafter\ifx\csname RF#3\endcsname\relax
               \global\advance\REFERENCENUMBER by 1
               \expandafter\xdef\csname RF#3\endcsname
                      {\the\REFERENCENUMBER}\fi
[\csname RF#1\endcsname,\csname RF#2\endcsname,\csname
RF#3\endcsname]\relax}
\def\refand#1#2{\expandafter\ifx\csname RF#1\endcsname\relax
               \global\advance\REFERENCENUMBER by 1
               \expandafter\xdef\csname RF#1\endcsname
                      {\the\REFERENCENUMBER}\fi
           \expandafter\ifx\csname RF#2\endcsname\relax
               \global\advance\REFERENCENUMBER by 1
               \expandafter\xdef\csname RF#2\endcsname
                      {\the\REFERENCENUMBER}\fi
            [\csname RF#1\endcsname,\csname RF#2\endcsname]\relax}
\def\Ref#1{\expandafter\ifx\csname RF#1\endcsname\relax
               \global\advance\REFERENCENUMBER by 1
               \expandafter\xdef\csname RF#1\endcsname
                      {\the\REFERENCENUMBER}\fi
             [\csname RF#1\endcsname]\relax}
\def\Refto#1#2{\expandafter\ifx\csname RF#1\endcsname\relax
               \global\advance\REFERENCENUMBER by 1
               \expandafter\xdef\csname RF#1\endcsname
                      {\the\REFERENCENUMBER}\fi
           \expandafter\ifx\csname RF#2\endcsname\relax
               \global\advance\REFERENCENUMBER by 1
               \expandafter\xdef\csname RF#2\endcsname
                      {\the\REFERENCENUMBER}\fi
            [\csname RF#1\endcsname--\csname RF#2\endcsname]\relax}
\def\Refand#1#2{\expandafter\ifx\csname RF#1\endcsname\relax
               \global\advance\REFERENCENUMBER by 1
               \expandafter\xdef\csname RF#1\endcsname
                      {\the\REFERENCENUMBER}\fi
           \expandafter\ifx\csname RF#2\endcsname\relax
               \global\advance\REFERENCENUMBER by 1
               \expandafter\xdef\csname RF#2\endcsname
                      {\the\REFERENCENUMBER}\fi
        [\csname RF#1\endcsname,\csname RF#2\endcsname]\relax}
\def\refadd#1{\expandafter\ifx\csname RF#1\endcsname\relax
               \global\advance\REFERENCENUMBER by 1
               \expandafter\xdef\csname RF#1\endcsname
                      {\the\REFERENCENUMBER}\fi \relax}

%

\def\NP{{ Nucl.\ Phys.\ }}
\def\PL{{ Phys.\ Lett.\ }}
\def\PR{{ Phys.\ Rev.\ }}

\def\PRL{{ Phys.\ Rev.\ Lett.\ }}
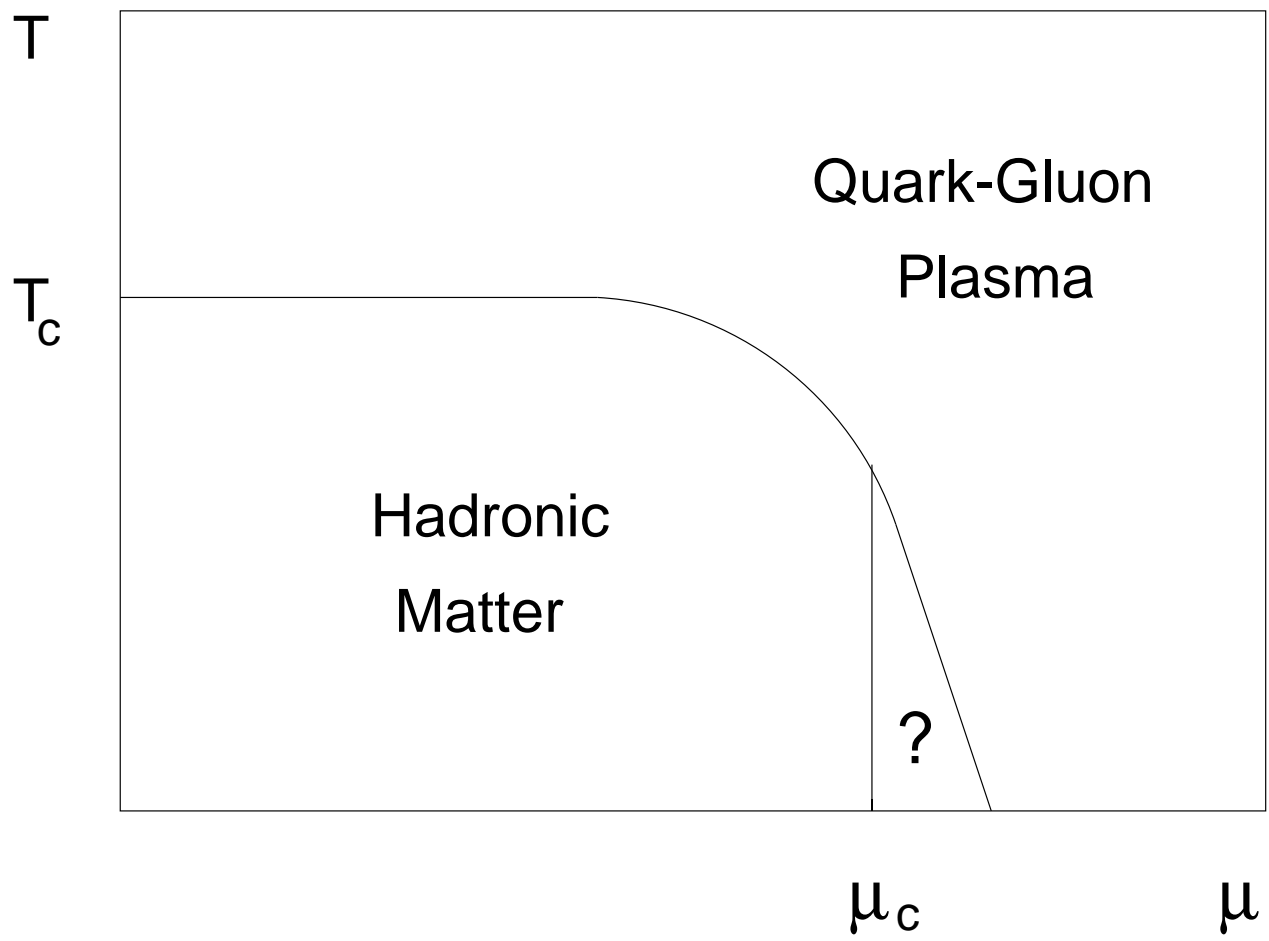
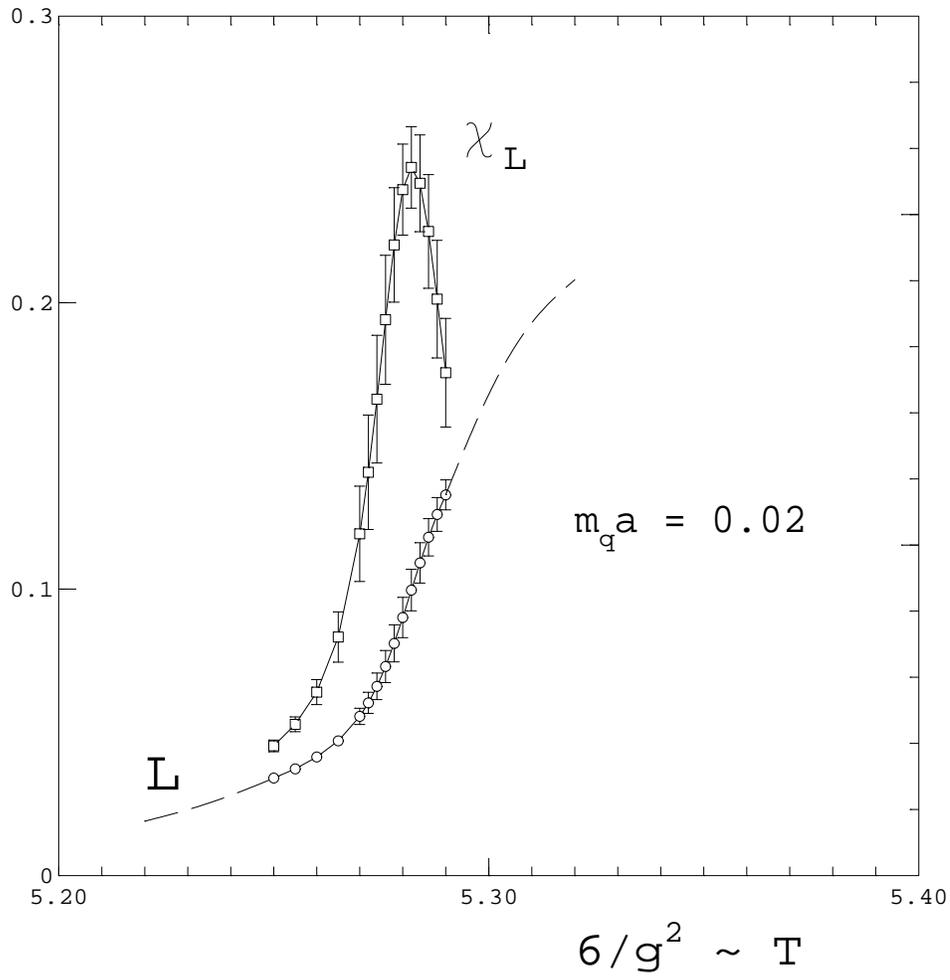
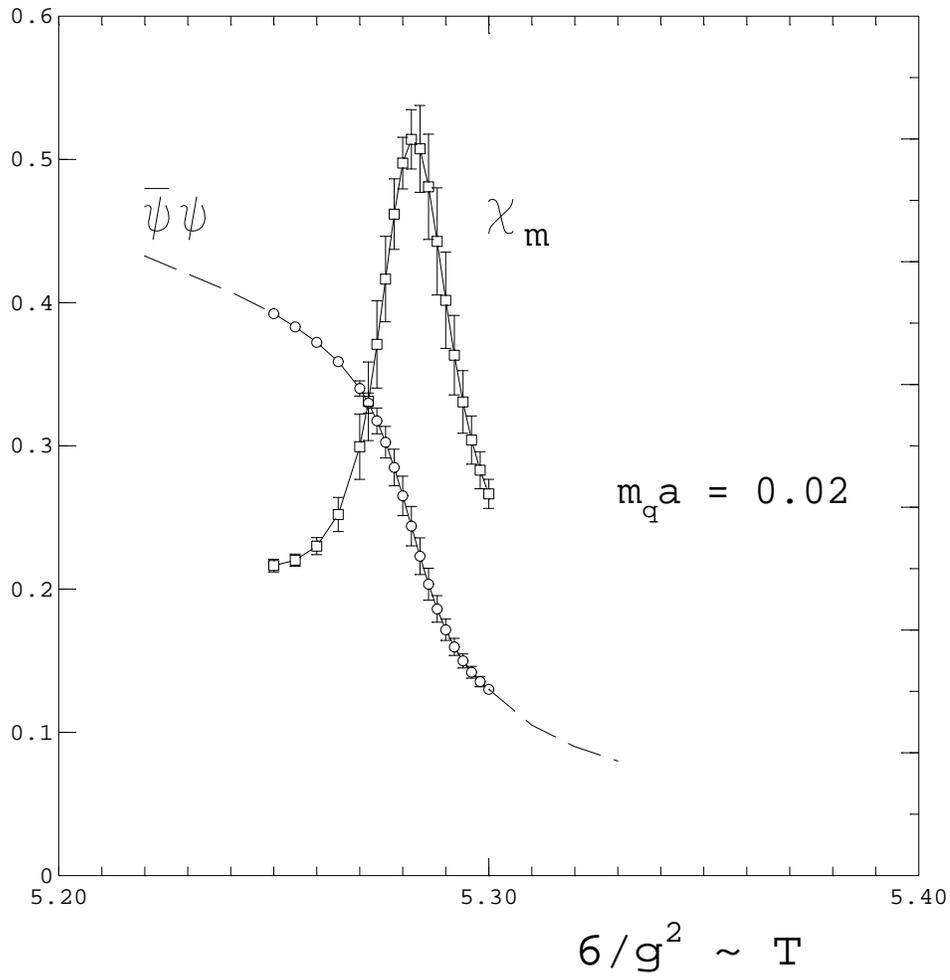
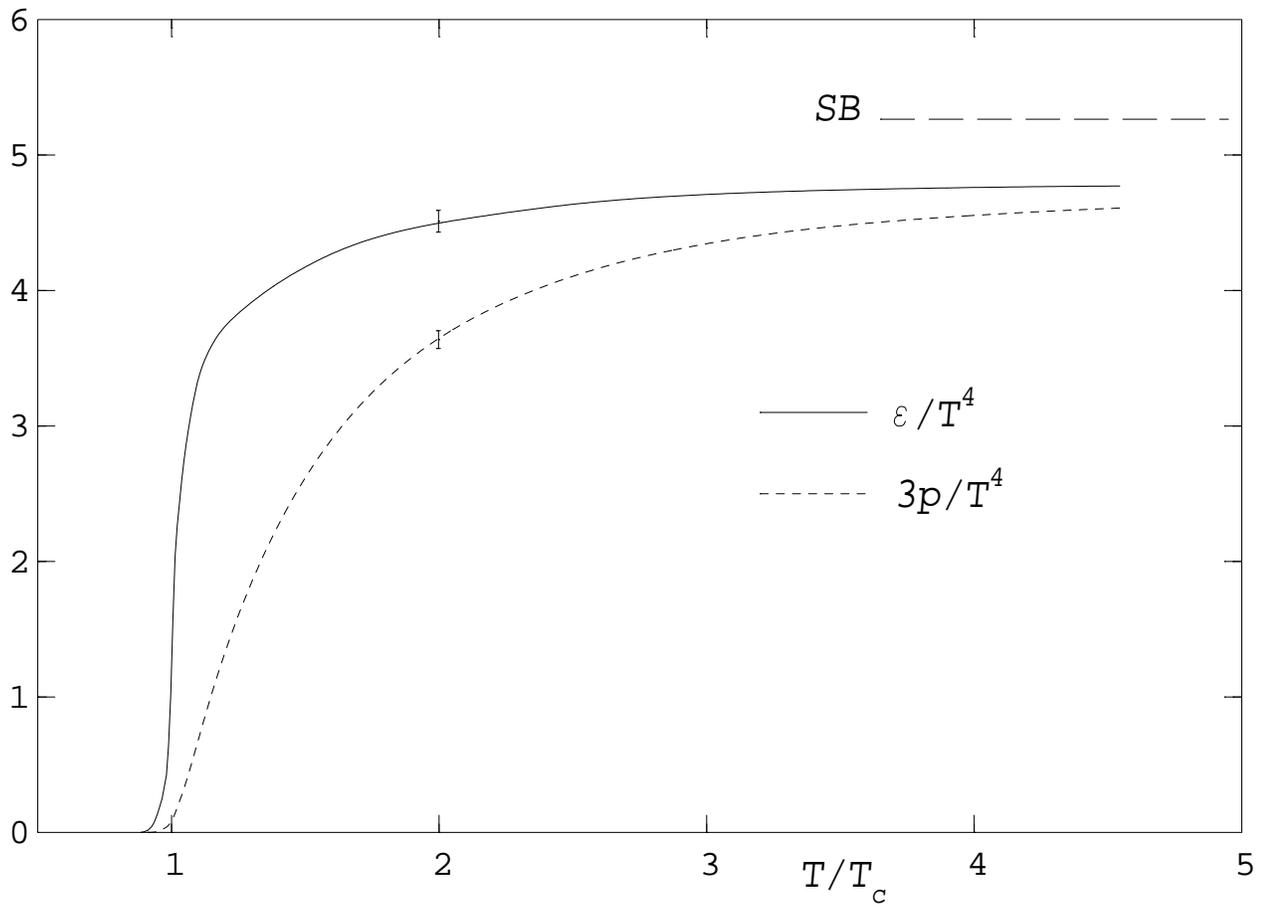
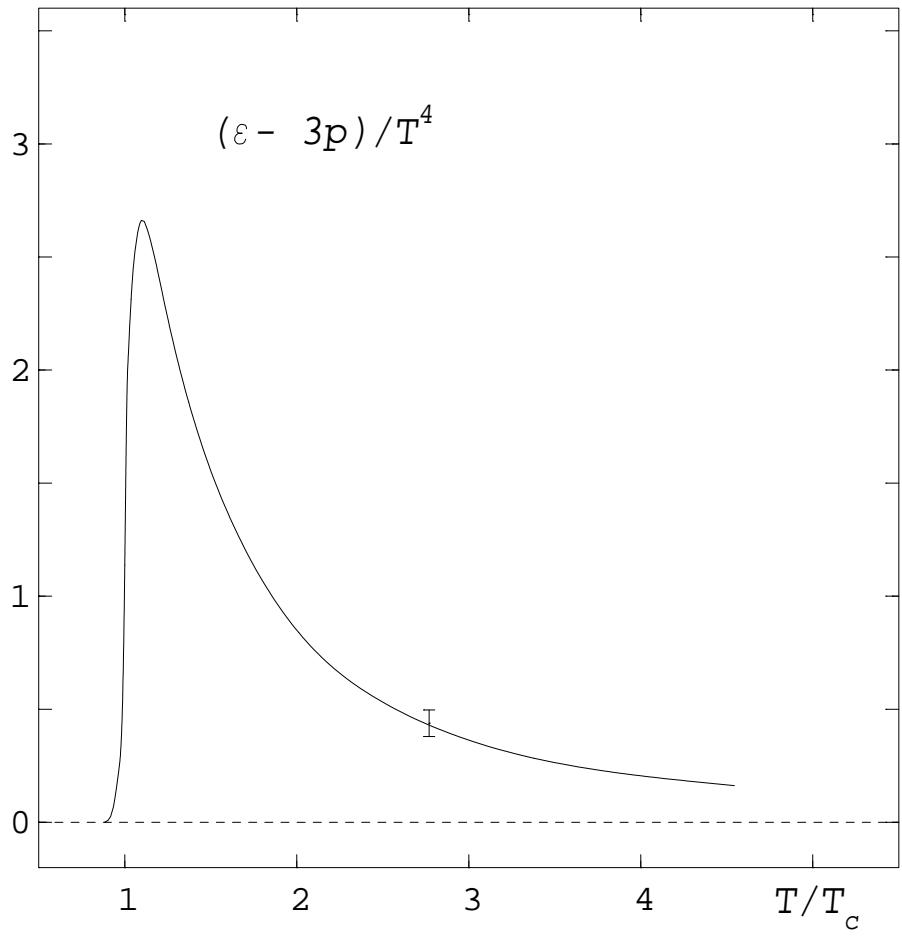
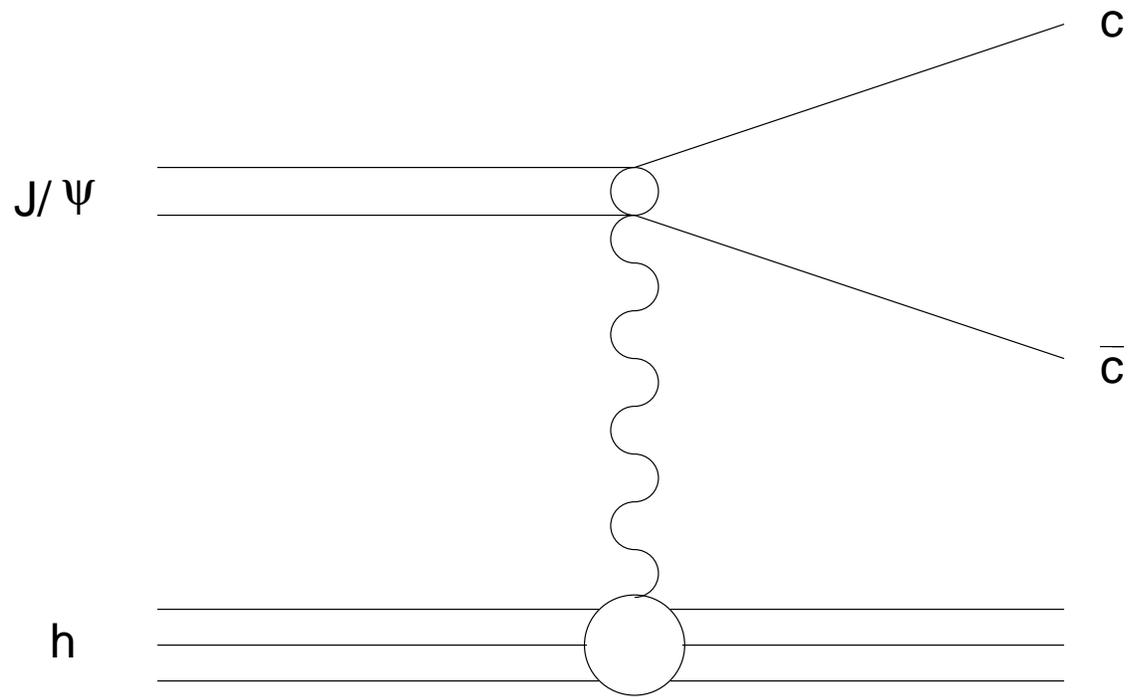
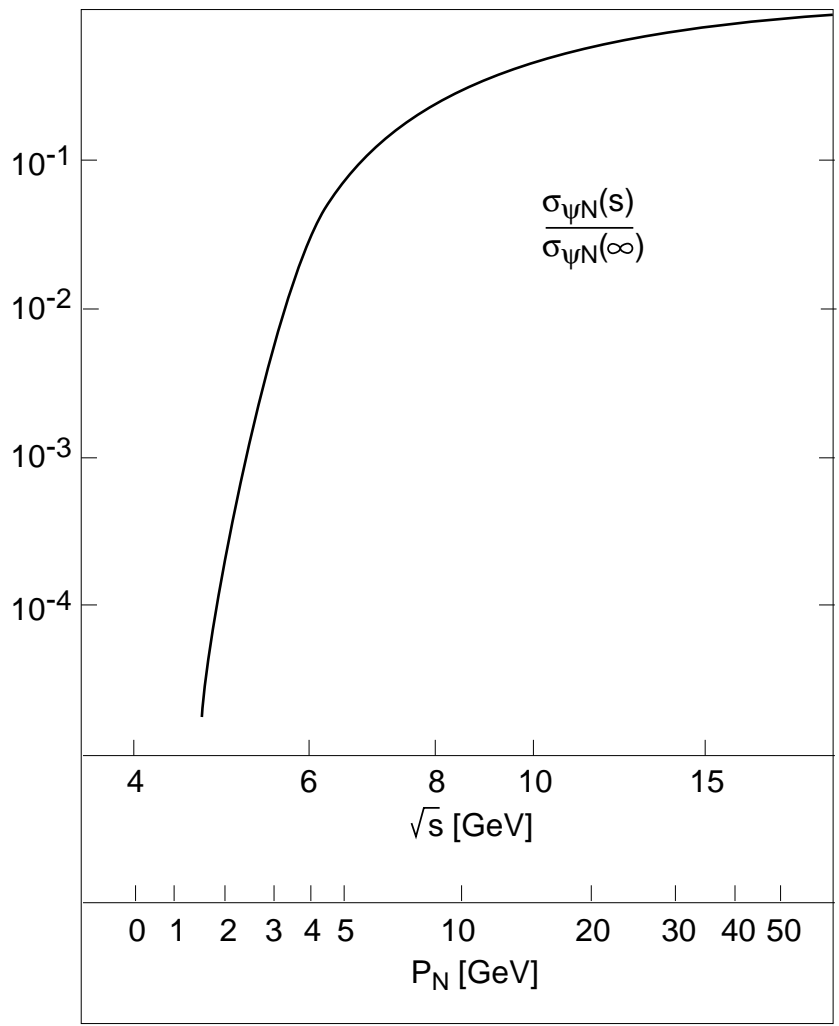
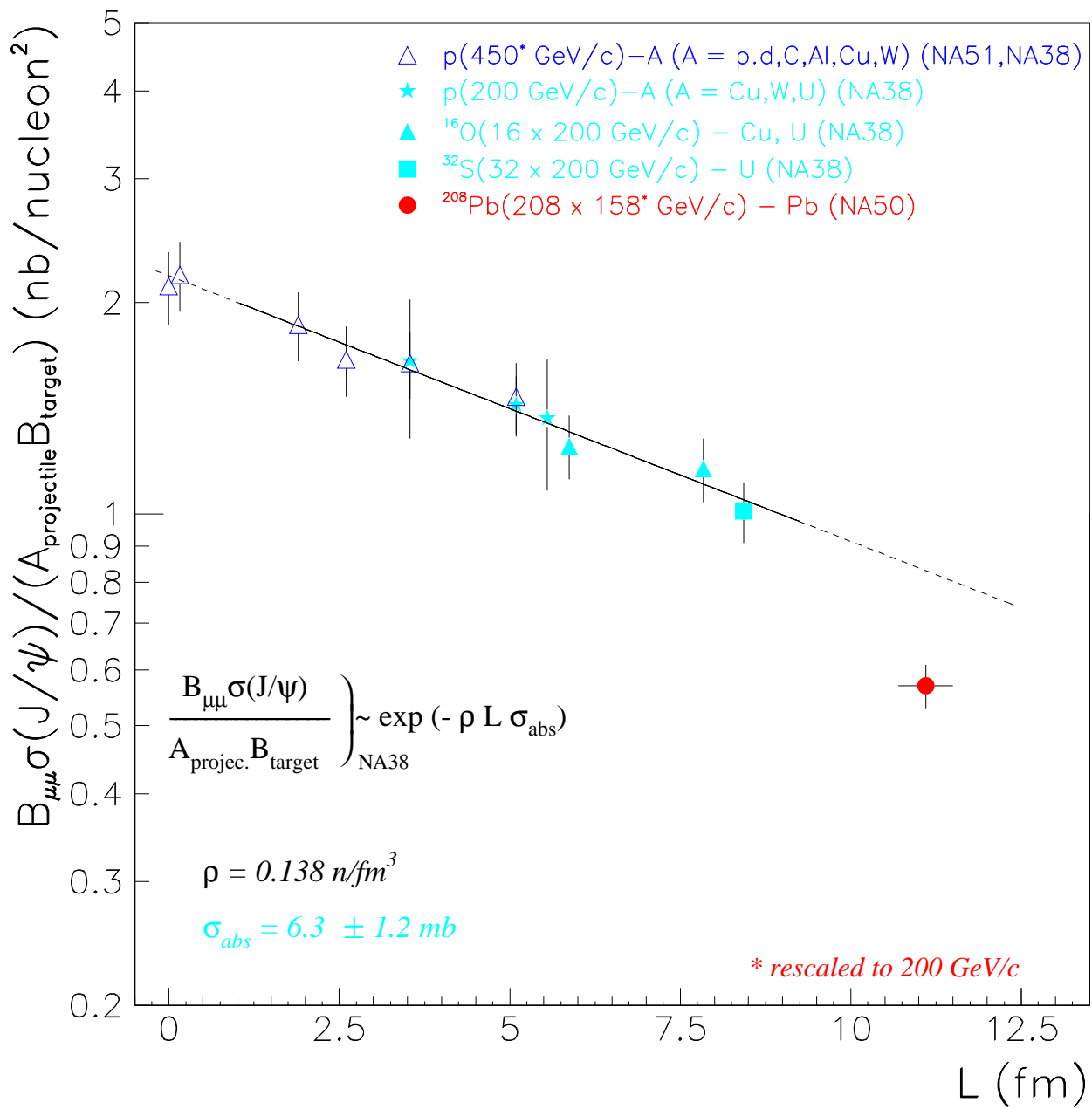
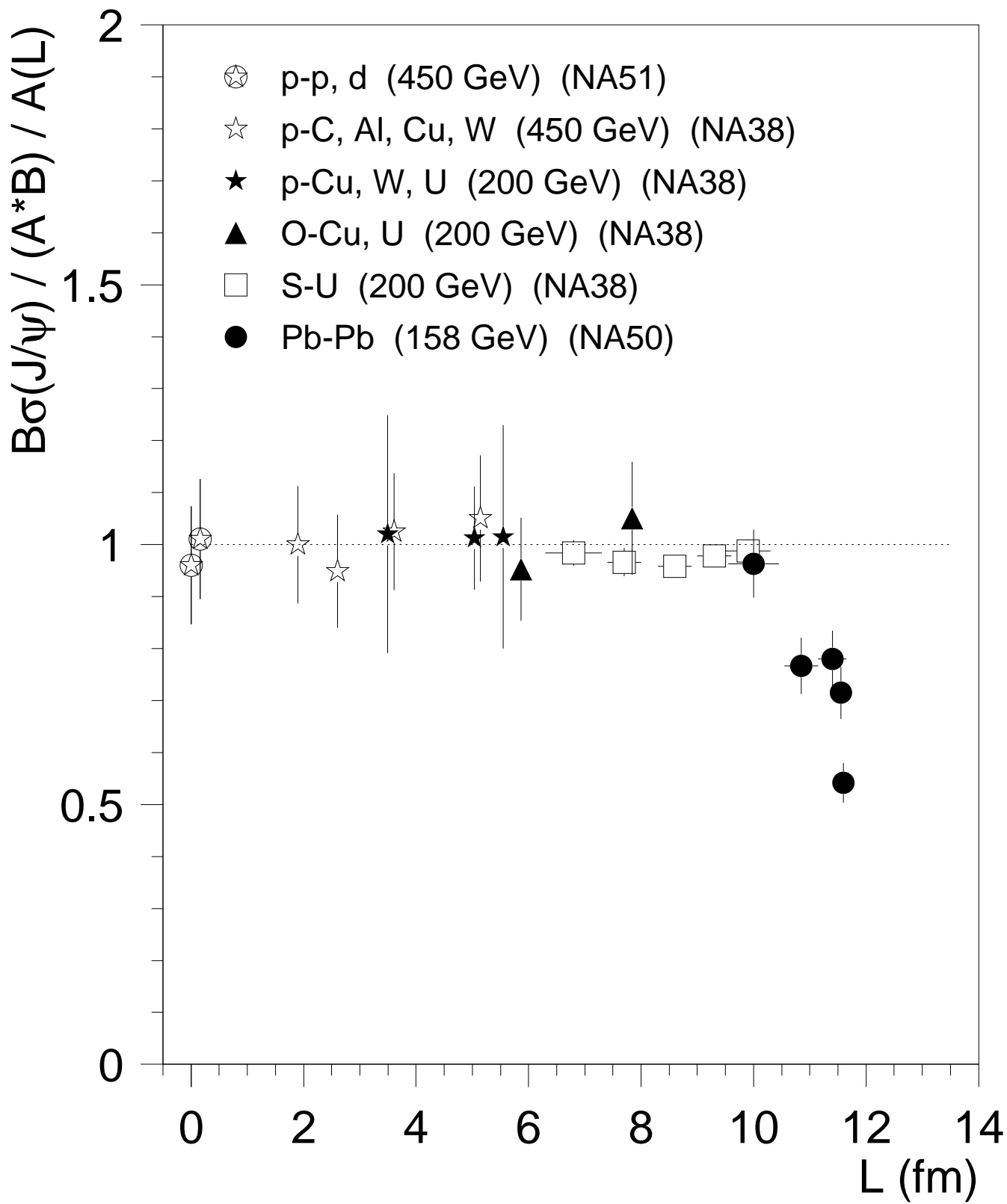
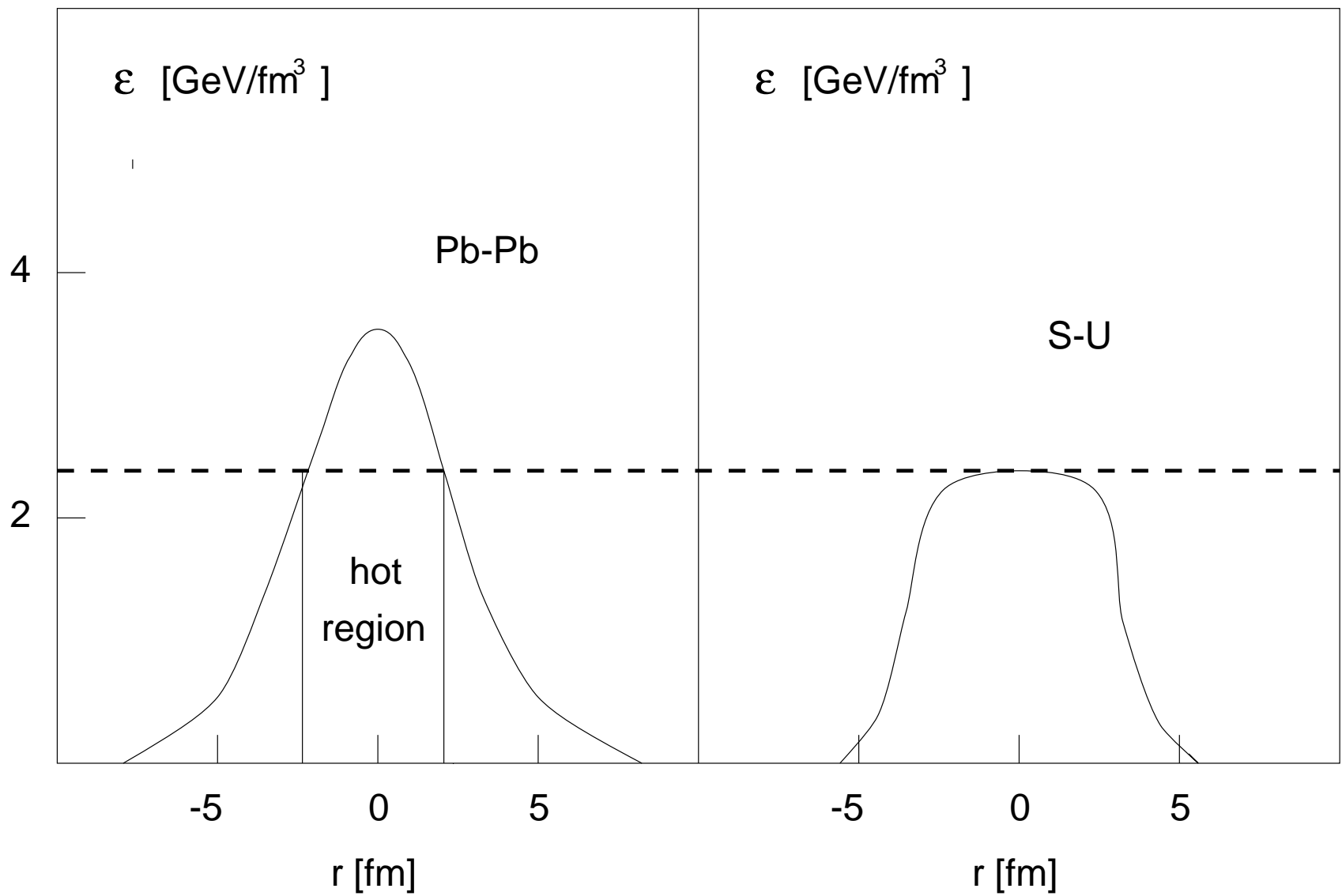
